\documentclass[twocolumn,amssymb, amsmath, nofootinbib, tightenlines,nobibnotes, aps, prl]{revtex4}
\usepackage{bm}%
\usepackage{graphicx}
\usepackage{amsfonts}

\begin{document}
\bibliographystyle{prsty}
\title{Spin lifetimes and strain-controlled spin precession of drifting electrons in zinc blende type semiconductors}
\author{M.~Beck}
%\email{beck@physik.uni-erlangen.de}
\author{C.~Metzner}
\author{S.~Malzer}
\author{G.~H.~D\"{o}hler}
\affiliation{Institut f\"{u}r Technische Physik I, Universit\"{a}t Erlangen--N\"{u}rnberg, Erlangen, Germany}
\date{\today}
\begin{abstract}
We study the transport of spin polarized electrons in n-GaAs using spatially resolved
continuous wave Faraday rotation. From the measured steady state distribution, we
determine spin relaxation times under drift conditions and, in the presence of strain,
the induced spin splitting from the observed spin precession. Controlled variation of
strain along [110] allows us to deduce the deformation potential causing this effect,
while strain along [100] has no effect. The electric field dependence of the spin
lifetime is explained quantitatively in terms of an increase of the electron temperature.
\end{abstract} \pacs{71.70.Fk, 72.25.Dc, 72.25.Rb, 85.75.-d}
\maketitle

Sufficiently long spin lifetimes and the possibility to manipulate the spin orientation
are required for the development of spintronics devices \cite{Zut04}. Electron spin
lifetimes  $\tau_s$ in semiconductors have been measured by means of the Hanle effect
(depolarization of photoluminescence in a magnetic field) \cite{Par69}, time resolved
photoluminescence \cite{Heb94}, and time resolved measurements of magneto-optical effects
(Faraday / Kerr rotation) \cite{Bau94}, which are suitable to measure $\tau_s$ in the
absence of holes. Although future spintronics applications are likely to depend on spin
{\em transport}, very little attention has been paid to the influence of an electric
drift field $F$ on $\tau_s$. While the possibility to transport the electron spin over
substantial distances in fields up to $F=100$~V/cm has been demonstrated several years
ago \cite{Kik99a}, first measurements of $\tau_s$ in this field range have only been
reported recently \cite{Kat04b}. Under the influence of strain, spin precession of the
drifting electrons \cite{Kik99a,Kat04,Cro04} has been observed and even the possibility
to generate spin polarized currents without magnetic materials or optical excitation has
been demonstrated \cite{Kat04b}. However, no theoretical interpretation of the observed
spin lifetimes and no quantitative analysis of the influence of strain on the observed
spin splitting has been given.

In this Letter, we present a method to study the lateral spin transport in thin films of
n-doped GaAs under steady state conditions, similar to the experiments reported recently
in \cite{Cro04}, but with the possibility to determine spin lifetimes and to quantify the
influence of strain. By the absorption of circularly polarized light we locally generate
a steady-state electron spin polarization and determine the spin drift length $L_s$ from
the spatial decay of spin polarization along the drift direction, which results from the
combined process of electron diffusion, drift and spin relaxation. As this signal can be
traced over several 100~$\mu$m while the signal varies by up to three orders of
magnitude, this cw method allows for accurate measurements of the electric field, doping
density and temperature dependence of $L_s$. Knowing the drift velocity $v_{\rm dr}$, the
spin relaxation time $\tau_s$ can be determined from $L_s$. In the presence of strain, we
observe spin precession in addition to the spatial decay. As realistic theoretical
predictions of the spin transport in an electric field were not available we have also
performed such calculations. We find excellent agreement between experiment and theory
both in the unstrained and the strained case and can deduce an experimental value for the
band structure parameter controlling the strain-induced spin precession.

\begin{figure}
\includegraphics[width=\linewidth]{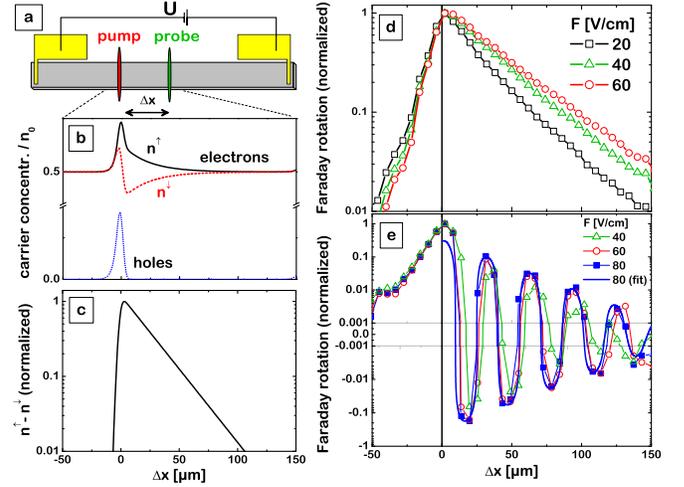}
\caption{\label{fig:principle} (a) Principle of the experiment; (b) example of an
expected spatial distribution of holes and spin up and spin down electrons (divided by
the equilibrium electron concentration $n_0$); (c) expected normalized electron spin
polarization; (d),(e): examples of the measured spatially resolved Faraday rotation in an
unstrained sample (d) and in a strained (here: unknown, uncontrolled strain) sample (e),
plotted on two logarithmic scales connected with a linearly scaled region in order to
visualize the exponential decay. The fit function to the data for $F=80$~V/cm is given by
$A\,\exp[(\Delta x-x_0)/\Lambda_d]\,\cos[2\pi(\Delta x-x_0)/\Lambda_p]$.}
\end{figure}
The principle of our measurement is illustrated in Fig.~\ref{fig:principle}(a--c). The
samples, grown by molecular beam epitaxy, are $1\times0.1$~mm$^2$ bars (oriented along
$[110]$, $[1\overline10]$, $[100]$, $[010]$) of silicon (n-type) doped (doping densities
$N_D = 10^{15}$~cm$^{-3}$\,\ldots\,$6\times10^{16}$~cm$^{-3}$) layers of GaAs (typically
1~$\mu$m thick; up to 5~$\mu$m for low $N_D$) in between two undoped
Al$_{0.3}$Ga$_{0.7}$As barriers (to avoid surface effects), lifted from the substrate by
epitaxial lift-off and attached to glass by van-der-Waals bonding. Circularly polarized
light from a cw laser diode (``pump''; typically 1.52~eV, 1~mW) excites spin polarized
electron hole pairs in the GaAs layer which diffuse and, in the case of a finite electric
field, drift laterally within the layer. The Faraday rotation of linearly polarized light
from a second cw laser diode (``probe''; typically 1.512~eV, 0.2~mW) is used to measure
the electron spin polarization.
After passing through the sample, located in a liquid helium flow cryostat (temperature
$T_{\rm L}\approx5$~K for all reported measurements), the probe light is split by a
Wollaston prism into two components, linearly polarized $\pm45^\circ$ with respect to the
initial polarization. The intensity difference of these components is detected using
lock-in technique at the sum of the modulation frequencies of the pump and probe laser
diodes. The difference between the signals obtained for left and right circular
excitation reflects the Faraday rotation induced by the spin polarized electrons. A
stepper motor varies the pump-probe distance along the direction of $F$.
The used edge emitting quantum well lasers give line-shaped spots on the sample
($\approx200\times5$~$\mu$m$^2$). Hence, since the layer thickness is much smaller than
the transport distances under consideration and the excitation is approximately
homogeneous across the sample, the transport can be treated as one-dimensional, which is
essential for the straightforward determination of $\tau_s$. Fig.~\ref{fig:principle}(b)
shows a simulation of the carrier concentrations along the drift direction for typical
parameters. Since the hole spin relaxes within picoseconds, the recombination ($\tau_{\rm
rec}\approx1$~ns) with electrons leaves behind a net electron spin polarization decaying
exponentially on the way towards the positive contact [Fig.~\ref{fig:principle}(b,c)]. In
unstrained samples, we observe this exponential decay along the drift direction
[Fig.~\ref{fig:principle}(d)]. While the ``downstream'' spin diffusion length $\Lambda_d$
along the drift direction increases with increasing field, the ``upstream'' diffusion in
the opposite direction, characterized by $\Lambda_u$ reduces. In strained samples, we
additionally observe spin precession along the transport direction
[Fig.~\ref{fig:principle}(e)], similar to the spatio-temporally resolved results reported
in \cite{Kik99a,Kat04,Kat04b} or the images of two-dimensional transport in \cite{Cro04}.
Since, as will shown below, the spin splitting depends linearly on the wave vector, the
precession length $\Lambda_p$ practically does not change with the electric field. The
thick solid line is a fit to the results for $F=80$~V/cm with the function
$A\,\exp[-(\Delta x-x_0)/\Lambda_d]\cos [2\pi(\Delta x-x_0)/\Lambda_p]$.

\begin{figure}
\includegraphics[width=\linewidth]{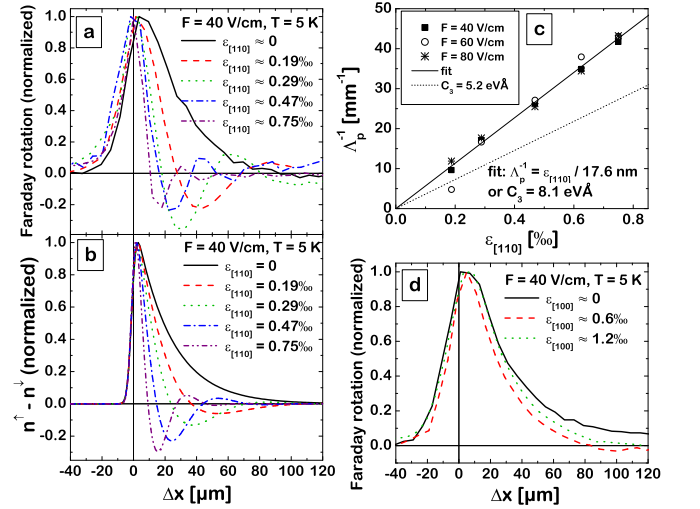}
\caption{\label{fig:strain} Spin transport at $F=40$~V/cm and
$N_D=1.4\times10^{16}\,$cm$^{-3}$ in strained samples. (a) measured Faraday rotation for
drift along $[1\overline10]$ and strain along $[110]$; (b) calculated spin polarization
with the parameters of (a) (using $C_3=5.2$~eV\AA); (c) $\Lambda_p^{-1}$ versus strain
$\epsilon_{[110]}$, linear fit corresponding to $C_3=8.1$~eV\AA, and expected result for
$C_3=5.2$~eV\AA\ (dotted line); (d) measured Faraday rotation for drift and strain along
$[100]$. }
\end{figure}

To quantify the influence of strain, we have carefully fabricated strain-free samples to
which we have then applied strain in a systematic manner. Unintentional uniaxial strain
during cool-down can be avoided by not attaching the glass substrate to any other
material. Biaxial in-plane strain should be small, due to the similar thermal expansion
coefficients of glass and GaAs. We control the strain by bending the glass substrate.
\begin{figure}[bht]
\includegraphics[width=0.66\linewidth]{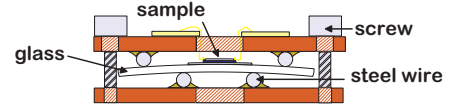}
\caption{\label{fig:strainsetup} Setup used to control the sample strain. Tightening the
screws essentially stretches the sample in the direction perpendicular to the wires
(uniaxial strain).}
\end{figure}
Fig.~\ref{fig:strainsetup} shows schematically our fixture used to vary the strain by
turning the screws. The magnitude of strain is determined from the focal length of light
reflected from the concave side of the glass. The measured spatial distributions of spin
polarization for drift along $[1\overline10]$ and strain along $[110]$ are shown in
Fig.~\ref{fig:strain}(a).
The theoretical expectation [Fig.~\ref{fig:strain}(b)] is obtained from the following
considerations. Strain lifts the spin degeneracy and, like an ``internal magnetic field''
yields spin precession. In zinc blende type semiconductors, the spin precession angular
frequency $\mathbf\Omega_p$ is given by \cite{Bir62,Dya82,Pik84,Pik88}
\begin{equation}\label{eq:Omega}
    \Omega_{p,x}=\frac{C_3}{\hbar} (\epsilon_{xy}k_y-\epsilon_{xz}k_z) \;\;\mbox{and cyclic permutations,}
\end{equation}
with the wave vector $\mathbf k$ and the strain tensor $\epsilon_{ij}$. Another
contribution including the diagonal elements of $\epsilon_{ij}$ exists, but is negligibly
small according to \cite{Pik84}. On the one hand, the statistical spin precession of
thermalized electrons leads to spin relaxation as a contribution to the D'yakonov-Perel'
(DP) mechanism. D'yakonov et al.\ \cite{Dya82} used the strain dependence of $\tau_s$ to
determine the constant $C_3=5.2$~eV\AA. For drifting electrons, on the other hand, i.e.\
if the average wave vector is non-zero, the average electron spin will in general
precess. For example if the drift is along [110] with velocity $v_{\rm dr}$, i.e.\ with
the average wave vector $\left<\mathbf k\right> = v_{\rm dr} m_e\hbar^{-1} \mathbf{\hat
e_{[110]}}$, $\mathbf{\hat e_{[110]}}$ being the unit vector along [110] and $m_e$ the
electron effective mass ($0.067m_0$ for GaAs), and if the uniaxial strain is along
$[110]$ or $[1\overline10]$ ($\epsilon_{xy} = \epsilon_{yx} = \pm
\epsilon_{[110],[1\overline10]}/2=:\epsilon/2$, with all other off-diagonal elements
vanishing) the average spin precession frequency vector, according to
Eq.~(\ref{eq:Omega}) will be $\left<\mathbf\Omega_p\right> = \epsilon C_3 v_{\rm dr}m_e
\hbar^{-2} \mathbf{\hat e_{[1\overline10]}}/2$.
Spin precession by $2\pi$ of the electrons generated with initial orientation along [001]
is expected at multiples of the precession period $\Lambda_p = 2\pi v_{\rm
dr}/\left|\left <\Omega_p\right >\right| = \frac{4\pi \hbar^2}{m_e \epsilon C_3} =
27.5$~nm$/\epsilon$ (with $C_3=5.2$~eV\AA). Thus, the observed precession neither depends
on the electric field nor on whether the strain is parallel or normal to the transport
direction. From the linear fit in Fig.~\ref{fig:strain}(c), we obtain $\Lambda_p =
17.6$~nm$/\epsilon$ or $C_3 = 8.1$~eV\AA. Although the scatter of this data is low, we
concede that the relatively coarse determination of $\epsilon$ may lead to a systematic
error of up to 30\%, which is the largest uncertainty. The absence of spin precession for
strain along $[100]$ [c.f.\ Fig~\ref{fig:strain}(d)] confirms that the influence of the
diagonal elements of $\epsilon_{ij}$ mentioned above is negligibly small.

We now focus on the influence of the drift field on electron spin relaxation. To our best
knowledge realistic calculations of spin transport and its dependencies on the applied
field, doping density, temperature and strain have not been reported before. Here we
sketch the fundamentals of our theory and the most important numerical results, while
details will be published elsewhere \cite{Bec05}. As Yu and Flatt\'{e} \cite{Yu02}
pointed out, electron spin transport can be treated in complete analogy to the ambipolar
transport of electrons and holes (e.g.\ \cite{Smi59}), with the simplification that here
``minority'' and ``majority'' carriers have the same properties, except for their spin.
Whereas the downstream and upstream diffusion lengths $\Lambda_d$ and $\Lambda_u$ exhibit
a somewhat complicated field dependence, their difference turns out to be the spin drift
length $L_s$, which is related to the spin lifetime by $L_s = \Lambda_d - \Lambda_u =
v_{\rm dr}\tau_s$. In the experiment, we determine $\tau_s$ from this relation, while
$v_{\rm dr}$ is obtained from the electron density, the sample geometry and the electric
current. For low fields, both $\Lambda_d$ and $\Lambda_u$ converge to the spin diffusion
length $\Lambda_s=\sqrt{D_e \tau_s}$, $D_e$ being the electron diffusion constant.

To calculate $\tau_s$, we basically follow the theory given in \cite{Pik84}. For the
fields, temperatures and doping densities in our experiments, the DP \cite{Dya72}
mechanism limited by ionized impurity scattering turns out to be the most important
contribution. For this mechanism, the spin relaxation rate for an electron with energy
$E$ is given by $\tilde\tau_s^{-1}=(32/105)\gamma_3^{-1}\tilde\tau_p\alpha_c^2 E^3
E_g^{-1} \hbar^{-2}$. Here, $\tilde\tau_p$ is the momentum relaxation time, $E_g$ the
band gap energy, $\alpha_c\approx0.063$, and $\gamma_3$ is a dimensionless parameter
taking into account the scattering cross section. D'yakonov and Perel' argued that for
small angle scattering (by ionized impurities) $\gamma_3\approx6$. However, a thorough
analysis shows that this is only true for extremely small scattering angles. We therefore
evaluate the parameter $\gamma_3$ numerically, using the scattering cross section
introduced by Ridley \cite{Rid77}, which reduces $\tau_s$ significantly (up to factor 5),
especially at low $T_{\rm L}$ and $N_D$.
\begin{figure}
\includegraphics[width=\linewidth]{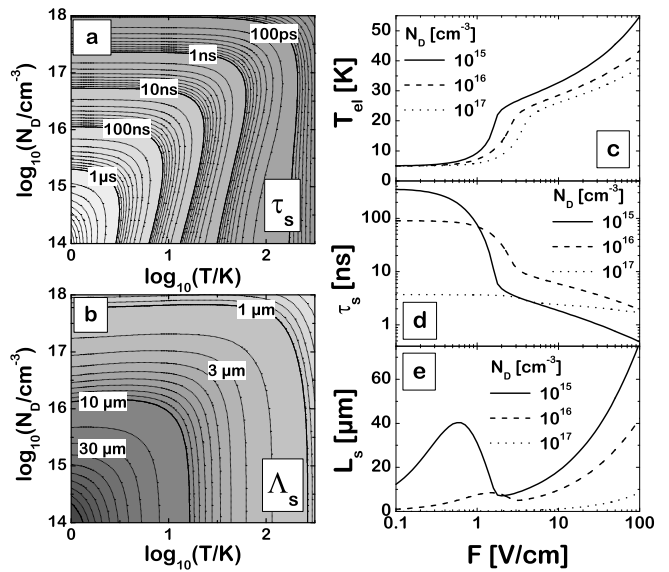}
\caption{\label{fig:taus} (a): Calculated spin relaxation times ($F=0$); (b): calculated
spin diffusion lengths ($F=0$); (c): expected field dependence of the electron
temperature at $T_{\rm L}=5$~K; (d): resulting spin lifetimes, (e): resulting spin drift
lengths.}
\end{figure}

The total spin relaxation rate $\tau_s^{-1}$ is obtained from the ensemble average of
$\left(\sum\tilde\tau_s\right)^{-1}$, where the sum is over all scattering processes.
This average has previously been taken either over a non-degenerate Boltzmann
distribution (e.g.\ \cite{Dya72}) or has been evaluated at the Fermi energy for a
strongly degenerate electron gas (e.g.\ \cite{Dzh02}). By using a Fermi distribution, we
cover both limiting cases and the intermediate regime. Taking into account also
scattering by acoustic phonons (deformation potential and piezoelectric) and polar
optical phonons, we obtain the spin relaxation times and the spin diffusion lengths at
$F=0$ shown in Figs.~\ref{fig:taus}(a) and (b), respectively. The calculated spin
relaxation times reproduce well the measured values \cite{Kik98a,Dzh02} for delocalized
electrons. However, as was pointed out by Kavokin \cite{Kav01} and Dzhioev et al.\
\cite{Dzh02}, this rate will not be correct for donor-bound electrons, i.e.\ at low
temperatures and doping densities. Surprisingly, the measured low temperature ($T_{\rm
L}=4.2$~K) spin lifetimes for $2\times10^{15}\,$cm$^{-3}< N_D <
2\times10^{16}\,$cm$^{-3}$, i.e.\ the range which is interpreted in terms of the
anisotropic exchange interaction in \cite{Dzh02}, can also be explained by the DP
mechanism in our calculations.

In the presence of an applied electric field $F$, the electron temperature $T_{\rm el}$
rises with respect to the lattice temperature $T_{\rm L}$ (e.g.\ \cite{Lun90}):
\begin{equation}
 T_{\rm el}=T_{\rm L}+\frac{2e}{3k_{\rm B}} v_{\rm dr} F
 \frac{\left<E\right>}{\left<E/\tau_E(E)\right>}\;.
\end{equation}
The angular brackets denote the averaging over the thermal distribution, and $\tau_E(E)$
is the energy relaxation rate. Hence, $T_{\rm el}$, $v_{\rm dr}$, $\tau_s$ and $L_s$, can
be calculated self-consistently for a given lattice temperature, doping density and
electric field from the energy and momentum relaxation rates. Figs.~\ref{fig:taus}(c--e)
show the expected dependencies of $T_{\rm el}$, $\tau_s$ and $L_s$ on the electric field
for several doping densities at $T_{\rm L}=5$~K. The steep increase of the electron
temperature at low $F$ to values of $T_{\rm el}\approx30$~K can be attributed to the fact
that all scattering mechanisms at very low electron temperatures are (qua\-si-)\-elastic
and therefore inefficient energy relaxation mechanisms. Only at an increased electron
temperature can the emission of optical phonons prevent a further increase of $T_{\rm
el}$. This strong increase of $T_{\rm el}$ is also reflected in $\tau_s$ and consequently
in $L_s$ (approximately: $\tau_s^{-1}\propto T_{\rm el}^3$).

\begin{figure}
\includegraphics[width=\linewidth]{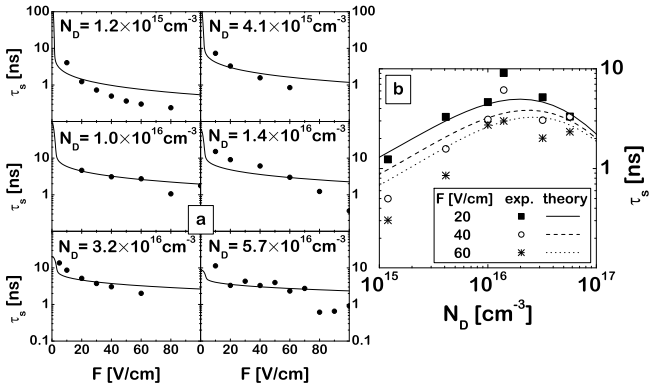}
\caption{\label{fig:tausexp} (a) Field dependence of the measured and calculated spin
lifetimes $\tau_s$ at $T_{\rm L}=5$~K and different doping densities $N_D$; (b) doping
density dependence at $F=20,40,60$~V/cm.}
\end{figure}
The field dependencies of the spin lifetimes at $T_{\rm L}=5$~K, extracted from
measurements like the one shown in Fig.~\ref{fig:principle}(d) are plotted in
Fig.~\ref{fig:tausexp}(a) together with the theoretical expectation, while the doping
density dependence can be seen in Fig.~\ref{fig:tausexp}(b). Over the whole range of
doping densities experiment and theory agree well at low fields. At all doping densities,
however, the experimental values decrease faster with increasing electric field than
expected, which may be due to sample heating by the electric current. Both our own
measurements and the field dependence of $\tau_s$ at low temperatures reported in
\cite{Kat04b} for $N_D=3\times10^{16}$~cm$^{-3}$ suggest that our calculations
overestimate the increase of $T_{\rm el}$ at small fields.

\begin{figure}
\includegraphics[width=0.7\linewidth]{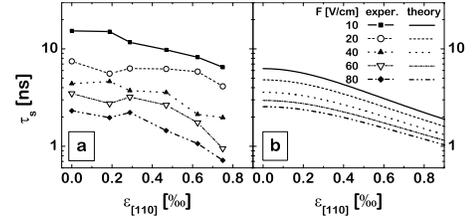}
\caption{\label{fig:tausvseps} Measured (a) and calculated (b) strain dependence of the spin
lifetime at different electric fields [same sample and geometry as in Fig.\
\ref{fig:strain}(a)].}
\end{figure}
Despite the stronger field dependence, the measured dependence of $\tau_s$ on uniaxial
strain along $[110]$ and drift along $[1\overline10]$ for $N_D=1.4\times10^{16}$~cm$^{-3}$, shown in Fig.~\ref{fig:tausvseps}(a),
agrees qualitatively with the theoretical expectation (b).

In conclusion, we have shown that spatially resolved Faraday rotation provides a
sensitive tool to investigate spin relaxation and spin precession. We have attributed the
observed strong decrease of the spin lifetimes at only small electric fields to the
increased electron temperature. The possibility to apply tunable uniaxial strain allowed
us to demonstrate the linear relation between the inverse spin precession length and the
strain applied along the $[110]$-directions, as well as the absence of spin precession
for strain along the $[100]$-directions. Thus, we were able to prove unambiguously that
the zero magnetic field spin precession is a consequence of the strain-induced spin
splitting first mentioned in \cite{Bir62} and to deduce the relevant band structure
parameter $C_3 = 8.1 \pm 2.5$~eV\AA.

\end{document}